\documentclass[twocolumn,aps, prb, showpacs,preprintnumbers,amsmath,amssymb]{revtex4}

\usepackage{graphicx}
\usepackage{dcolumn}
\usepackage{bm}

\begin{document}
\newcommand{\Arg}[1]{\mbox{Arg}\left[#1\right]}
\newcommand{\bb}{\mathbf}
\newcommand{\braopket}[3]{\left \langle #1\right| \hat #2 \left|#3 \right \rangle}
\newcommand{\braket}[2]{\langle #1|#2\rangle}
\newcommand{\be}{\[}
\newcommand{\br}{\vspace{4mm}}
\newcommand{\bra}[1]{\langle #1|}
\newcommand{\braketbraket}[4]{\langle #1|#2\rangle\langle #3|#4\rangle}
\newcommand{\braop}[2]{\langle #1| \hat #2}
\newcommand{\dd}[1]{ \! \! \!  \mbox{d}#1\ }
\newcommand{\DD}[2]{\frac{\! \! \! \mbox d}{\mbox d #1}#2}
\renewcommand{\det}[1]{\mbox{det}\left(#1\right)}
\newcommand{\ee}{\]} 
\newcommand{\eg}{\textbf{\\  Example: \ \ \ }}
\newcommand{\Imag}[1]{\mbox{Im}\left(#1\right)}
\newcommand{\ket}[1]{|#1\rangle}
\newcommand{\ketbra}[2]{|#1\rangle \langle #2|}
\newcommand{\kp}{\arccos(\frac{\omega - \epsilon}{2t})}
\newcommand{\ldos}{\mbox{L.D.O.S.}}
\renewcommand{\log}[1]{\mbox{log}\left(#1\right)}
\newcommand{\Log}{\mbox{log}}
\newcommand{\Modsq}[1]{\left| #1\right|^2}
\newcommand{\nb}{\textbf{Note: \ \ \ }}
\newcommand{\op}[1]{\hat {#1}}
\newcommand{\opket}[2]{\hat #1 | #2 \rangle}
\newcommand{\occ}{\mbox{Occ. Num.}}
\newcommand{\Real}[1]{\mbox{Re}\left(#1\right)}
\newcommand{\so}{\Rightarrow}
\newcommand{\sol}{\textbf{Solution: \ \ \ }}
\newcommand{\thetafn}[1]{\  \! \theta \left(#1\right)}
\newcommand{\tin}{\int_{-\infty}^{+\infty}\! \! \!\!\!\!\!}
\newcommand{\Tr}[1]{\mbox{Tr}\left(#1\right)}
\newcommand{\kb}{k_B}
\newcommand{\rad}{\mbox{ rad}}
\preprint{APS/123-QED}

\title{Impurity segregation in graphene nanoribbons}

\author{S. R. Power$^{(a)}$, V. M. de Menezes$^{(a,b)}$, S. B. Fagan$^{(c)}$ and M. S. Ferreira$^{(a)}$}
\email{ferreirm@tcd.ie}
\affiliation{%
(a) School of Physics, Trinity College Dublin, Dublin 2, Ireland \\
(b) Departamento de F\'{\i}sica, Universidade Federal de Santa Maria, UFSM, 97105-900, RS, Brazil \\
(c) \'Area de Ci\^encias Tecnol\'ogicas, Centro Universit\'ario Franciscano, UNIFRA, 97010-032, Santa Maria - RS, Brazil
}%

\date{\today}

\begin{abstract}
The electronic properties of low-dimensional materials can be engineered by doping, but in the case of graphene nanoribbons (GNR) the proximity of two symmetry-breaking edges introduces an additional dependence on the location of an impurity across the width of the ribbon. This introduces energetically favorable locations for impurities, leading to a degree of spatial segregation in the impurity concentration. We develop a simple model to calculate the change in energy of a GNR system with an arbitrary impurity as that impurity is moved across the ribbon and validate its findings by comparison with \emph{ab initio} calculations. Although our results agree with previous works predicting the dominance of edge disorder in GNR, we argue that the distribution of adsorbed impurities across a ribbon may be controllable by external factors, namely an applied electric field. We propose that this control over impurity segregation may allow manipulation and fine-tuning of the magnetic and transport properties of GNRs.
\end{abstract}

                 
\maketitle

Low-dimensional carbon materials such as fullerenes and nanotubes have been in the scientific limelight for the past two decades. Research initially instigated by their peculiar physical properties has been further motivated by their potential as future components of nano-electronic devices. Intensive research for such an extended period of time has inevitably led to a number of advances and byproducts, one of which is the experimental production of graphene \cite{riseofgraphene, neto:graphrmp, Hiura:2004exfoliate, Novoselov:2005graphene, Zhang:graphenehall, Berger:2006epitaxial, Berger:2004epitaxial}. Composed of a single sheet of hexagonally bonded carbon atoms, graphene can be further manipulated to produce narrow-width stripes commonly referred to as graphene nanoribbons (GNRs). As a matter of fact, GNRs of various widths and geometries can be experimentally realized by cutting mechanically exfoliated graphene sheets \cite{Hiura:2004exfoliate, Novoselov:2005graphene, Zhang:graphenehall}, or by patterning graphene grown epitaxially \cite{Berger:2006epitaxial, Berger:2004epitaxial}. 

Because doping is one effective way of tailoring the electronic properties of a material, it is worth investigating how a GNR is affected by the introduction of impurities. A crucial difference to the bulk system is the existence of two symmetry-breaking edges, which are expected to make some of the physical properties of the GNR dependent on the impurity position. Although previous studies have investigated how the conductance of GNRs \cite{biel:ribbondoping, rigo:Nidopedribbons, mucciolo:graphenetransportgaps} depends on the location of impurities, one crucial aspect that seems to have been overlooked is that this dependence arises also in the energetics of the doping process. In other words, the binding energy of a dopant depends on its position across the ribbon. As a result, we can identify energetically favorable locations for impurities, leading to some degree of spatial segregation in the impurity concentration. Bearing in mind that impurity segregation is known to occur at symmetry-breaking interfaces between two materials due to quantum interference effects \cite{Sutton:1995interfaces, Castro:2004segregation}, it should come as no surprise that the proximity of the two edges of a GNR is capable of inducing similar segregational features in the impurity distribution. What is surprising in the case of GNR is that the segregation may be easily controllable by external factors, which opens the road to manipulating the impurity distribution within a ribbon. We argue that this might be a possible route to engineering some of the physical properties of GNRs.  

\begin{figure}
\includegraphics[height = 2.85cm] {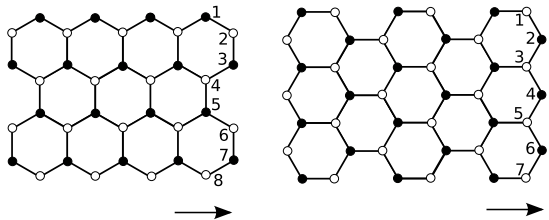}
\caption{Schematic drawings of GNR. Left (right) panel shows a small cross section of a 4-ZGNR (7-AGNR). Filled and hollow symbols represent the two distinct sub-lattices of a hexagonal structure. On each panel, numbered sites indicate the positions where impurities will be included either substitutionally or as an adatom. The arrows indicate the periodicity direction.}
\label{figure_1}
\end{figure}

To account for the position dependence of the binding energy we must define the geometry of the host ribbon and the type of impurity to be introduced. We consider a GNR that is of infinite length but has a finite width. Two possible edge geometries are considered, namely zigzag (ZGNR) and armchair (AGNR) edged ribbons, schematically depicted on the left and right panels of Figure \ref{figure_1}, respectively. An integer preceding the GNR abbreviations refers to the number of zigzag chains (or half the number of atoms) across the width of a ZGNR or the number of atoms across an AGNR. For instance, the left panel of Figure \ref{figure_1} shows a small cross section of a 4-ZGNR where the numbered sites label the positions within the GNR for clarity. Similarly for the 7-AGNR shown on the right panel. Both panels show sites marked as filled or hollow circles representing atoms from each of the two distinct, intersecting sub-lattices of the hexagonal graphene atomic structure. These sublattices are non-equivalent in the case of ZGNRs. We assume the impurity to take the form of a single atom that may either adsorb to the surface of the GNR or replace a host atom in the lattice. These are referred to as adatoms or substitutional impurities, respectively. 

The electronic structures of graphene related materials in general are known to be well described using a nearest-neighbour tight-binding Hamiltonian of the form 
\begin{equation}
{\hat h}_r = \sum_{\ell,\ell^\prime} |\ell\rangle \,\gamma_{\ell,\ell^\prime} \, \langle \ell^\prime | \,\,,
\label{nntb}
\end{equation}
where $|\ell\rangle$ labels a $\pi$-orbital centred at the site $\ell$,  $\gamma_{\ell,\ell^\prime} \equiv \gamma = -2.7\mathrm{eV}$ is the nearest-neighbour electronic hopping in graphene and $\ell^\prime$ is summed over the nearest neighbours of $\ell$. This simple Hamiltonian provides a good first approximation to the band structure of GNRs and will be used throughout this work, although further considerations \cite{Fujita:zigzagedgestates, Son:halfmetallic, Son:ribbonenergygaps, White:thirdNN, Cresti:disorderedgraphenereview} are required to more closely replicate the results of \emph{ab initio} calculations. As far as the atomic impurity is concerned, it is important to distinguish between adatom and substitutional impurities. Adatoms can be concisely expressed by the Hamiltonian ${\hat h}_a = \sum_i | i \rangle \epsilon_i \langle i |$, where $ | i \rangle$ represents the atomic orbital associated with the level $\epsilon_i$. For the sake of simplicity, we choose to represent the electronic structure of the atomic impurity by a single atomic orbital $| a \rangle$, making the sum over $i$ dispensable. We must also account for the interaction between the ribbon and the adatom. An adatom can connect in a number of ways to the host ribbon. We shall consider here the simplest, or ``top'' configuration, where the adatom is assumed to connect to only a single carbon atom. Other possibilities include the ``bridge'' and ``hollow'' configurations, where the adatom sits midway between two carbon atoms and connects to both or where it sits above the centre of a hexagon and connects to the surrounding six carbon atoms, respectively. The results for these more complex arrangements do not differ greatly from those for the ``top'' configuration, which we account for here with a connecting potential ${\hat V}_a = |a\rangle \, t \, \langle j | +  |j \rangle \, t^* \, \langle a |$ . The index $j$ labels the GNR atomic site that is in closest contact with the impurity atom, and $t$ describes the hopping parameters between lattice and impurity orbitals. Strictly speaking, a correction to the on-site potential associated with the state $| j \rangle$ should be included \cite{claudia:onsitecorrection} but this is not done here as it does not affect the key features of our results.  For substitutional impurities, the Hamiltonian structure is even simpler. In this case the introduction of an impurity can be accounted for by the following potential ${\hat V}_s = |j \rangle \, \delta \, \langle j |$, where $\delta$ is a correction to the on-site potential at site $j$ reflecting the different electrostatic characteristics of the inserted impurity. 

The quantity of interest is the difference between the total energies of two distinct configurations: one in which GNR and impurity are connected and another in which they are far apart. This can be summarized by evaluating the total energy variation due to the perturbation  ${\hat V}_a$ (${\hat V}_s$) for the case of adatom (substitutional) impurities. One can write the total energy of a system as the electronic structure contribution added to a repulsive energy term \cite{Cohen:totalenergy}, in which the latter has been given a formal correspondence with modern density functional theory (DFT) \cite{Foulkes:TB&DFT}. This latter contribution, not expected to carry a strong position dependence, should play only a minor role in the segregation features. Therefore, the band-structure contribution to the total energy variation becomes the most relevant quantity to be calculated and is given by the so-called Lloyd Formula\cite{Lloyd}

\begin{equation}
 \Delta E = \frac{1}{\pi}\, \mathrm{Im} \int_{-\infty}^{E_F} \mathrm{d}E \,\mathrm{ln} \left\lbrace \mathrm{det} \left( \hat{1} - \hat{\cal G}(E) \hat{V} \right) \right\rbrace \,\,,
\label{lloydgeneral}
\end{equation}
where $\hat{V}$ is the perturbation potential describing the considered impurity, $\hat{\cal G}$ is the single-particle Green function operator associated with the perturbation-free Hamiltonian and $E_F$ is the Fermi energy of the system. As $\hat{V}$ is very sparse, the only nonzero element of $\hat{\cal G}$ is $\langle j | \hat{\cal G} | j \rangle$, which happens to be the only position-dependent element in Eq.(\ref{lloydgeneral}). 

The segregation is now studied by selecting the type of impurity and its position within the GNR, calculating the matrix element of $\hat{\cal G}$ and finally evaluating the integral in Eq.(\ref{lloydgeneral}). Some numerical care is required to solve this integral. We take advantage of the fact that $\hat{\cal G}$ is analytic in the upper half of the complex energy plane and use an integration contour along the imaginary axis. Numerically, this is far more efficient since it avoids the van Hove singularities that exist along the real axis. 

\begin{figure}
\includegraphics[width = 8cm] {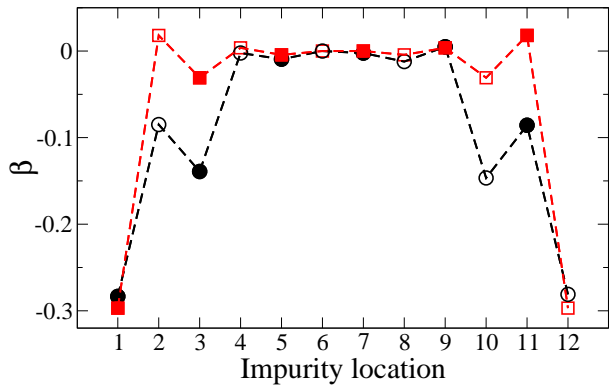}
\caption{Segregation function $\beta$ for substitutional impurities on different locations of a 6-ZGNR. Red squares indicate the results for the model calculations; black circles those for DFT calculations for Ti atoms. Hollow and filled symbols indicate which sub-lattice contains the substitutional replacement. Lines are guide to the eyes only. }
\label{figure_2}
\end{figure}

We consider first the case of substitutional impurities in a 6-ZGNR. It is appropriate to analyze the position dependence of the total energy through a renormalized energy scale that simplifies the comparison between distinct cases. To this end we define the segregation energy function (SEF) $\beta \equiv (\Delta E - \Delta E_c)/|\Delta E_c|$, where $\Delta E_c$ is the electronic contribution to the total energy variation evaluated at the centre of the GNR and which is taken as a reference energy. This dimensionless quantity describes the percentage deviation of the energy variation with respect to its value at the central position. The square symbols of Fig. \ref{figure_2} represent the values of $\beta$ for all positions across the width of a 6-ZGNR with substitutional impurities ($\delta = \gamma$) and points to a scenario in which they prefer to occupy the edges of the GNR with an energy variation that is predicted to be 30\% lower than at the center. This preference for edge sites is also true for adatoms and has already been reported by previous authors \cite{biel:ribbondoping, rigo:Nidopedribbons, li:ribbonedgedefects}. What is remarkable in our results is the way in which $\beta$ varies when the impurity position moves to the central region of the GNR. Rather than simply vanishing, it does so in a non-monotonic fashion pointing to the existence of a few local minima separating the lowest value at the edges from the central zero. 

To test whether such a non-monotonicity in the position dependence of the binding energy could be an artifact of our simple model, we carried out DFT calculations in which a similar 6-ZGNR was substitutionally doped with Ti atoms located at different positions across the ribbon. 
These calculations were carried out using the SIESTA \cite{SIESTA:2002} code with a 98-atom supercell. Double zeta basis set plus polarization functions were employed and the exchange-correlation function was adjusted using the generalised gradient approximation according to the parameterization proposed by Perdew, Burke and Ernzenhof\cite{PBE}. To represent the charge density, a cutoff of 170 Ry for the grid integration in real space was used. The interactions between the ionic cores and the valence electrons were described with normconserving Troullier-Martins pseudopotentials\cite{troullier}. The structural optimizations were performed with the conjugate gradient approximation\cite{SIESTA:2002} until the residual forces were smaller than 0.05 eV/\AA. 

The results of these calculations are shown by the circular symbols in Fig. \ref{figure_2}  and display  similar behavior for $\beta$ as those from our simple model, shown by square symbols, including excellent agreement at the ribbon edges. The existence of local minima was also reproduced at the same locations, albeit with slightly different values for $\beta$.  Such an excellent agreement with DFT results reassures us that our simple model contains the essential ingredients to describe the effect of impurity segregation in GNRs. With this model we can consider ribbons of all sizes and geometries as well as include an arbitrary number of impurities, if necessary. 

\begin{figure}
\includegraphics[width =  \columnwidth] {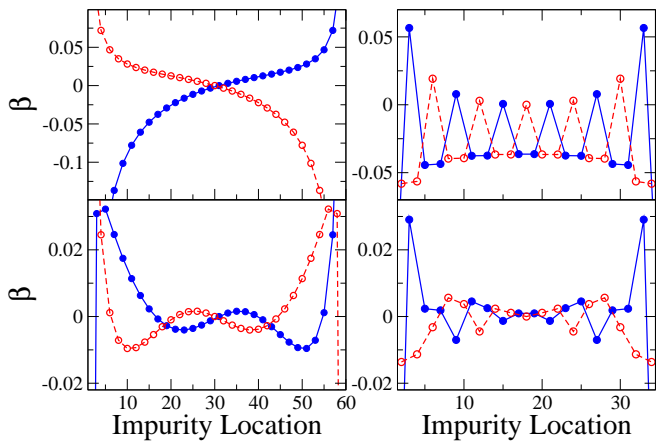}
\caption{Segregation function $\beta$ for adsorbed atoms on a 30-ZGNR (left panels) and 35-AGNR (right panels) for two values of the Fermi Energy : $E_F = 0.0t$ (top panels) and $E_F = 0.2t$ (bottom panels). The filled (blue) and unfilled (red) circles represent adsorption sites from the two distinct sublattices. The solid and dashed lines connect sites within a given sublattice. Here we have focused on the central region of both ribbons, but the reader should note that the edge sites, not shown here, are the most favourable adsorption sites}
\label{figure_3}
\end{figure}

A point worth raising is that the location of substitutional impurities usually follows the existence of defects and vacancies, often induced by ionic irradiation \cite{Nordlund:ionirrad, Chung:boroncarbon, krashirrad}. In this scenario, impurities will occupy the sites surrounding the defects, which means that edge-induced impurity segregation will play a minor role in the doping process. However, for adsorbed atoms the situation is very different. In this case the impurities will adsorb at the most energetically convenient sites. Thus the position dependence of the binding energy is a key factor in determining where the impurities will be adsorbed. As previously anticipated, there is very little qualitative difference in our model between the substitutional and adsorbed cases, which suggests similar non-monotonic variations in the segregation function across the ribbon. This is shown in Fig. \ref{figure_3} where the SEF for adsorbed impurities in the central region of a 30-ZGNR (35-AGNR) is displayed on the left (right) panels. Filled (blue) and hollow (red) symbols indicate above which sub-lattice the impurities are located. The top left shows that the segregation function for ZGNR alternates between positive and negative depending on which sub-lattice the impurity is above, similar to the case for substitutional impurities. There is a clear distinction between the filled and hollow points, in the sense that on the left half of the ribbon the former are energetically more favorable as adsorption sites for the impurities, whereas the latter becomes preferable on the right half of the GNR. A solid (dashed) line linking the values of $\beta$ for hollow (filled) sites is also shown. Both lines intersect at the center of the GNR, where $\beta=0$, confirming that the preferential location for impurities changes from one sub-lattice to another precisely at this location. Similar non-monotonicities in the SEF are also found for AGNR, shown on the right panels of Fig. \ref{figure_3}, although in this case there is no obvious distinction between the two sub-lattices in regard to the most energetically favorable position. For clarity, Fig. \ref{figure_3} focuses on the central regions of the ribbons, but in both cases the impurities are found to attach much more readily to edge atoms (not shown here) than to central atoms. The edge value of $|\beta|$ is much larger in the zigzag case, which can be reconciled with the existence of localized edge states at the (half-filling) Fermi energy in these ribbons\cite{Fujita:zigzagedgestates}. The sublattice dependent non-monotonicity disappears if we consider the ``bridge'' or ``hollow'' configurations, as the adatoms connect to carbon atoms from both sublattices and the effect is averaged out. However, a marked preference for edge sites with a decay towards the centre, as seen here for the ``top'' configuration, is still present.

As in the case of substitutional impurities, we performed DFT calculations for adsorbed Ti atoms on a 6-ZGNR.
It was found that on each side of the ribbon one of the sublattices was dominant. When an adatom was released above a site belonging to this sublattice it would remain there. However, adatoms released over sites from the other sublattice tended to migrate either to sites above the dominant sublattice or to more complex intermediary sites. The other sublattice was found to assume the dominant role on the opposite side of the ribbon. The migration behaviour described makes it difficult to make a direct comparison with the simple model SEF, as we did for the substitutional case. However, the existence of this type of behaviour confirms qualitatively the results of our simple model, which predicts sites from a single sublattice to be favoured on either side of the ribbon, as seen in the top left panel of Fig \ref{figure_3}.  Once again, the agreement between the results based on our simple model and those obtained by DFT calculations are encouraging and suggest that this model can be used to shed some light in situations where {\it ab-initio} calculations are unable to do so. 

The ease with which the Fermi level, $E_F$, of graphene-based structures can be manipulated with external gate voltages\cite{riseofgraphene} adds an extra ingredient to the study of impurity segregation in GNR. The bottom panels of Fig. \ref{figure_3} shows the SEF for both zigzag and armchair ribbons when the Fermi energy is shifted away from half-filling by a mere $3\%$ of the graphene bandwidth. The solid and dashed lines used to distinguish between the two different sub-lattices are clearly modified as $E_F$ is changed. Whereas the AGNR remains without any clear favorites for the most energetically preferred locations, there is a striking effect on ZGNR. In this case the two lines intersect not one but five times indicating that the energetically favorable location for the adsorption of impurities changes periodically between the two sub-lattices forming a striped pattern across the ribbon width. This oscillatory feature is also present for the ``bridge'' and ``hollow'' configurations.

It is important to note the general nature of the model for the SEF we have constructed and used in this work. We have made no assumptions about the atomic species used as the impurity. Although it is possible to fit our tight-binding parameters to DFT calculations, this is not necessary to recover the qualitative features of the results shown above. Indeed, our results for substitutional atoms with arbitrary tight-binding parameters match the results of a full \emph{ab initio} calculation for Ti atoms to a  high degree of accuracy (Fig \ref{figure_2}). This suggests that the non-monotonic behaviour of $\beta$ displayed in the above results is independent of the impurity species chosen, and depends only on underlying graphene lattice and how the impurity is embedded into it. This is evident from the form of Eq. (\ref{lloydgeneral}), where the position dependence arises solely in the Green function matrix element of the host ribbon. Therefore similar behaviour can be expected if the impurities considered possess a magnetic moment. Recent works have established that a long range magnetic coupling can exist between magnetic atoms embedded in graphene-related materials\cite{AntonioDavidIEC,David:IEC,David:PhD,DynamicNJP}. Furthermore, it is found that certain magnetic dopants adsorbed onto sites within the same sublattice prefer to align ferromagnetically, whereas those on opposite sublattices prefer an antiferromagnetic alignment\cite{DavidSpinValve, brey:graphenerkky, saremi:graphenerkky, yazyev:graphenemagnetism2, santos:Nidopedgraphene}. Thus, if in a given region of a ribbon a majority of the magnetic dopants adsorb onto one of the sublattices, it follows that these dopants may prefer to align ferromagnetically, resulting in a net magnetic moment in this region. Similarly, a  net magnetic moment with opposite sign should form in regions where the other sublattice is preferential. By controlling the Fermi energy, it may therefore be possible to manipulate the width of magnetic domains across the ribbon. In this manner, it may be possible to engineer doped GNRs with magnetic properties determined by the application of an electric field during the impurity adsorption phase.

\begin{figure}
\includegraphics[width = 0.45\textwidth] {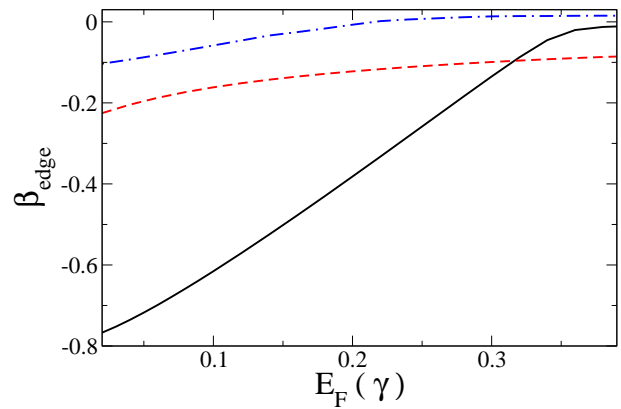}
\caption{The segregation function at the edge of a ribbon, $\beta_{edge}$, measures how favourable the edge site of a ribbon is as an adsorption site relative to the central site. When $E_F=0$, the edge site is far more favourable for both a 10-ZGNR (black, solid line) and 11-AGNR (red, dashed line). However by shifting $E_F$, the edge and central sites become more equally favourable, particularly in the case of ZGNRs. Also shown is the case of an adatom in the ``hollow'' configuration in a 30-ZGNR (blue, dot-dashed line). In this case the edge and central sites correspond to atoms adsorbed in the middle of a hexagon at the edge or at the centre of the nanoribbon. }
\label{figure_4}
\end{figure}

The transport properties of a graphene nanoribbon have been shown to be dependent on the position of a single doped impurity\cite{biel:ribbondoping, rigo:Nidopedribbons}. The introduction of an impurity in general introduces quasibound states into the band structure of GNRs. These in turn lead to the formation of dips or gaps in the conductance of these ribbons at energies corresponding to the quasibound states\cite{li:ribbonedgedefects}. The position of these single impurities across the ribbon width is found to affect the energies at which these conductance dips occur, as well as their width and depth. 
Due to the preference of impurities to locate themselves at the edges of a ribbon, much of the work examining extended disorder in GNRs has focused exclusively on edge disorder\cite{li:ribbonedgedefects, evaldsson:ribbonedgeanderson}. However, recent work\cite{mucciolo:graphenetransportgaps} has compared the effects of edge disorder in GNRs to those of bulk disorder, where impurities are allowed to distribute uniformly throughout the ribbon. A marked difference has been found between these two cases. For example, mild edge disorder produces only a small effect in the conductance of ZGNRs, whereas bulk disorder can lead to a more dramatic suppression of the conductance, with roughly the opposite effect observed for AGNRs.
This difference between edge and bulk disorder suggests that controlling the impurity distribution across a ribbon may be a viable method of engineering its transport properties. 
Fig. \ref{figure_4} shows $\beta_{edge}$, the value of $\beta$ at the edge of a ribbon, as a function of $E_F$ for a 10-ZGNR and 11-AGNR, and also for a ``hollow'' type adatom on a 30-ZGNR. When this quantity approaches zero, the edge and central sites are equally favourable. We see from Fig. \ref{figure_4} that as $E_F$ is increased from half-filling, for ZGNRs at least, the preference for adsorption at edge sites is decreased continually until edge and central sites are almost equivalent. This suggests it may be possible to engineer ribbons with the transport properties associated with edge disorder, bulk disorder or any intermediate position on the continuum between these two.
This presents itself as a possible method for fine-tuning the resistance properties of a ribbon device.

In summary, we have demonstrated that the energy variation when an impurity is introduced into a GNR exhibits non-monotonic behaviour as a function of the location of the impurity. This results in a degree of spatial segregation in the impurity distribution across a GNR. In the case of ZGNRs, the non-monotonicity is connected to the sublattices of the graphene atomic structure. Furthermore, we found that the qualitative features of this result are indepedent of the specific impurity type, and depend only on the properties of the underlying graphene host. A simple theoretical model for calculating how the energy variation changes across a ribbon has been developed and is in excellent agreement with the results of DFT calculations for both substitutional and adsorbed impurities. We postulated that control of the adsorbed impurity segregation within a ribbon is possible by adjusting the Fermi energy. We thus argued that, due to the sublattice dependence of magnetic interactions and the defect position dependence of transport within graphene, the magnetic profile and electronic properties of a GNR may be engineered by exploiting this control of the impurity segregation.

\begin{acknowledgments}
The authors acknowledge support received from Science Foundation Ireland and the Irish Research Council for Science, Engineering and Technology under the EMBARK initiative. Computational resources were provided by the HEA IITAC project managed by the Trinity Centre for High Performance Computing.
\end{acknowledgments}




\end{document}